\documentclass[journal=apchd5,manuscript=article,email=false]{achemso}

\usepackage{chemformula} 
\usepackage[T1]{fontenc} 

\author{Maxime Duchet}
\author{Sorin Perisanu}
\author{Stephen T. Purcell}
\author{Eric Constant}
\author{Vincent Loriot}
\affiliation[ILM]
{Univ Lyon, Univ Claude Bernard Lyon 1, CNRS, Institut Lumi\`ere Mati\`ere, F-69622, VILLEURBANNE, France.}
\author{Hirofumi Yanagisawa}
\author{Matthias F. Kling}
\affiliation[Second University]
{Department of Physics, Ludwig-Maximilians-Universit\"at Munich, Am Coulombwall 1, 85748 Garching, Germany.}
\alsoaffiliation[Institute]
{Max Planck Institute of Quantum Optics. Hans-Kopfermann-Stra\ss e 1, 85748 Garching, Germany}
\author{Franck Lepine}
\author{Anthony Ayari}
\affiliation[ILM]
{Univ Lyon, Univ Claude Bernard Lyon 1, CNRS, Institut Lumi\`ere Mati\`ere, F-69622, VILLEURBANNE, France.}

\title{Femtosecond laser induced resonant tunneling in an individual quantum dot attached to a nanotip}


\begin{document}

\begin{tocentry}

\includegraphics[width=3.5in]{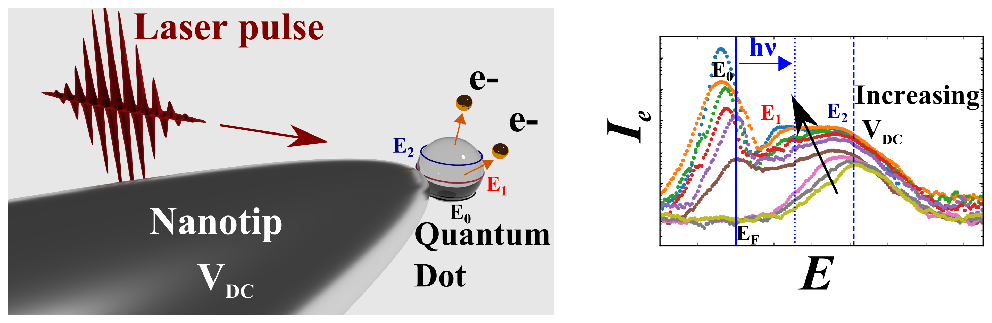}

\end{tocentry}

\begin{abstract}
  Quantized nano-objects offer a myriad of exciting possibilities for manipulating electrons and light that impact photonics, nanoelectronics and quantum information. In this context, ultrashort laser pulses combined with nanotips and field emission have permitted to renew nano-characterization and control electron dynamics with unprecedented space and time resolution reaching femtosecond and even attosecond regimes. A crucial missing step in these experiments is that no signature of quantized energy levels has yet been observed. We combine \textit{in situ} nanostructuration of nanotips and ultrashort laser pulse excitation to induce multiphoton excitation and electron emission from a single quantized nano-object attached at the apex of a metal nanotip. Femtosecond induced tunneling through well-defined localized confinement states that are tunable in energy is demonstrated. This paves the way for the development of ultrafast manipulation of electron emission from isolated nano-objects including stereographically fixed individual molecules and high brightness, ultra-fast, coherent single electron sources for quantum optics experiments.
\end{abstract}

Recent experimental developments in electron sources using metallic nanometric tips and ultrashort laser pulses have given a new impulse to nano-characterization instruments such as time resolved scanning tunneling\cite{yoshioka2016real,cocker2016tracking}, scanning near-field-optical\cite{huber2016ultrafast}, point projection\cite{quinonez2013femtosecond} or transmission electron microscopes\cite{feist2015quantum,houdellier2018development} by improving the observation of spatiotemporal processes at the subnanometric and subfemtosecond scale or in the quantum regime\cite{kruit2016designs}. These sources and instruments give complementary information to well-established ultrafast electron diffraction techniques on homogeneous materials and thin films\cite{zewail2010four,miller2014femtosecond} but more importantly open new scientific pathways into the exploration and exploitation of quantum systems.

Ultrafast field emission sources driven by coherent laser pulses play a central role in unraveling the pertinent high field physics and achieving ultimate source characteristics. The interaction between a nanotip and an intense laser pulse has led to the observation of multiphoton emission, above threshold ionization and electron rescattering plateaus in the energy distribution of the emitted electrons \cite{yanagisawa2011energy,kruger2011attosecond,herink2012field,kruger2018attosecond}. Until now the experiments were performed for field emitters with a continuum electron energy density. However, the large number of electron energy levels involved in these promising demonstrations of ultrafast phenomena should limit coherent control strategies and their interpretation in general. Consequently, it can be expected that working with smaller emitters, preferentially minimally interacting with a support tip, will strengthen quantum mechanical effects and open new avenues in electron manipulation.

Downsizing metallic emitters has already been pushed almost to its limit. Remarkable results were obtained on ultra sharp emitter down to 10 nm radius for tungsten\cite{kruger2011attosecond} (W) and gold\cite{herink2012field} tips. An experiment on a "single atomic emitter", was even briefly mentioned in ref.~\citenum{hommelhoff2009extreme}, but no significant change compared to larger tips was observed. 
Studying the photoemission of an isolated object with well-separated energy states\cite{hofer1997time,reutzel2019coherent} requires to attach nano-objects with an electronic structure that is preserved from strong interaction with the metallic tip. The interaction between an ultrashort laser pulse and a plasmonic nanostructure has attracted much attention in the ultrafast community, allowing to study individual nano-objects deposited on a surface\cite{grubisic2012plasmonic,dombi2013ultrafast,marsell2015nanoscale,racz2017measurement} and recently attempts to study femtosecond field emission on individual free standing nano-objects on tips have emerged for instance on gold nanorods\cite{ahn2017attosecond}, carbon nanotubes\cite{li2017quiver} and diamond nano-needles\cite{torresin2019conduction,borz2019photoassisted} but no signature of quantized energy levels has yet been reported. In Ref.~\citenum{gruber2013spectral} a quantum dot was coupled to a metallic nanowire but fluorescence emission was studied and not photoemission.

For field emitters, a quantum dot (\textit{i.e.} a nanometric object weakly coupled to the electron reservoir of the tip via a tunnel barrier and showing discrete electron energy levels) can be easily fabricated by \textit{in situ} nanostructuration \cite{binh1992field}. Although the exact chemical composition of these quantum dots is still an open question (see Supporting Information I.G.), it appears that their properties do not depend on the tip material nor its exact fabrication method. They all show electron energy spectrum with individual energy peaks below the Fermi energy and specific electric field dependence (see below). However, their energy levels above the Fermi energy has never been probed experimentally and these quantum dots were never studied under laser illumination. In this article, we propose to use such individual quantum dots for field assisted photo-emission in order to show femtosecond laser induced resonant tunneling through quantized energy levels. This new process opens opportunities in the study of ultrafast electron dynamics in individual nano-objects.

\begin{figure}
  \includegraphics[width=13cm]{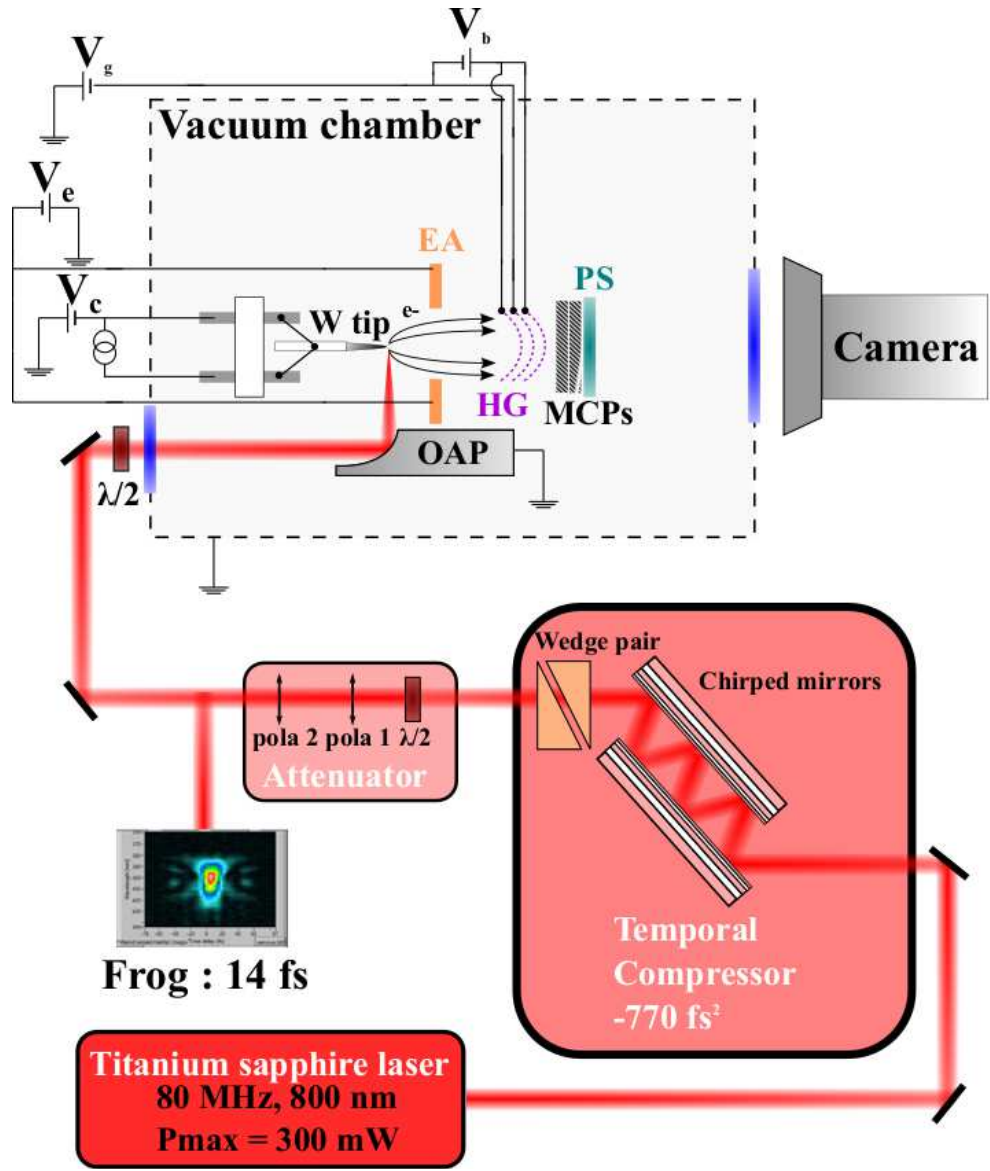}
  \caption{\label{set} Schematic of the ultra-fast beam-line (OAP : Off-Axis Parabolic Mirror, pola : polarizer, $\lambda/2$ : half-wave plate, FROG : Frequency-resolved optical gating) and field emission set-up in an ultra high vacuum chamber (V$_c$ : cathode voltage, V$_e$ extraction voltage, EA : Extraction anode, MCPs : microchannel plates, PS : phosphor screen) for electron energy analysis with retarding field hemispherical grids (HG) with grid voltage V$_g$ and a bias voltage V$_b$ (see Supporting Information I.E).}
\end{figure}

\section*{In situ nanostructuration and field emission characterization of the nano-object}

The experimental system illustrated in Figure \ref{set} is a standard field emission set-up \cite{good1956field,yanagisawa2011energy,kruger2011attosecond,herink2012field} with an ultra-high vacuum chamber ($5\times10^{-10}$ Torr) an electrochemical etched $<$111$>$ tungsten tip cathode at a negative voltage, an extraction anode at a positive voltage and a retarding field analyzer. The DC voltage reported below is the voltage difference between the anode and the tip. A Ti:sapphire laser oscillator (80 MHz repetition rate, a  photon energy $h\nu = $1.55 eV, a pulse duration of 14 fs and a 5 $\mu$m waist) can be focused to the apex of the tungsten nanotip. We captured the emitted electrons with a two-stage microchannel plate, a phosphor screen and a camera to get the spatial distribution, and a retarding field analyzer with lock-in detection to characterize the electron energy distributions (see Supporting Information I). \textit{In situ} nanostructuration was performed in order to grow in ultra-high vacuum conditions an individual nano-object at the apex of the tip. This growth method is a two steps process similar to methods developed in the past \cite{PhysRev.119.85,binh1992electron,nagaoka2001field,binh1992field,ming1996new} and is described in detail in Supporting Information I.G. The first step leads to the formation of a ”buildup” tip and the  second consists  in  emitting electrons at 100 nA current for 10 minutes.

\begin{figure}
  \includegraphics[width=8cm]{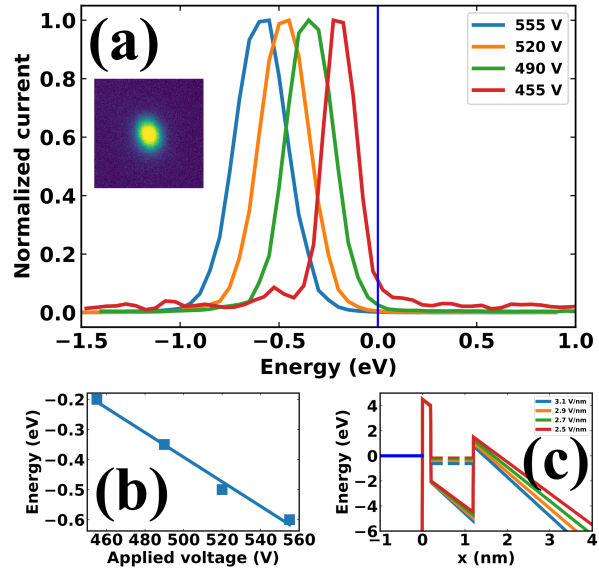}
  \caption{\label{fig1} (a) Experimental energy spectrum of the emitted electrons from a nanostructure grown on a W$<$111$>$ tip for different applied DC voltages. The vertical line represents the position of the Fermi energy inside the tip. Left inset: field emission pattern of a nanostructure. (b) Energy shift of the maximum of the energy peak in (a) as a function of the applied voltage. The line is a fit of the experimental data. (c) Calculated energy band diagram of the quantum dot attached to a metallic tip for different applied DC voltages. The horizontal line represents the position of $E_F$ inside the tip. The horizontal dashed lines represent the position of the emitting level in the quantum dot for different applied DC voltages.}
\end{figure}

Figure \ref{fig1}a presents the electron energy spectra and the emission patterns after the nano-object growth. The standard three fold rotational symmetry pattern of a clean W$<$111$>$ tip has been replaced by a single spot at the center of the screen. The applied DC voltage required to have a given current is twice smaller for the nano-object. The electron energy spectra obtained after growth consist of a peak with a maximum at an energy significantly lower than the Fermi energy ($E_F$) of the tip and this peak shifts down linearly with the applied voltage as shown in Figure \ref{fig1}b. This behavior is notably different from the voltage dependence of the spectra obtained in the case of metallic nanotips (see Supporting Information I.H. for the pattern and energy spectra of the W tip before the nano-object growth).

In field emission a linear shift of the energy peak with the voltage is a signature of an electron emission process from quantized energy levels that is described by a resonant tunneling model\cite{duke1967field,gadzuk1970resonance}. Resonant tunneling through a single nano-object on a field emission tip has been experimentally demonstrated for deposited molecules\cite{muller1950rendering}, atoms\cite{clark1968field}, clusters\cite{lin1991observation} or as here for quantum dots \textit{in situ} nanostructured on sharp metallic tips\cite{binh1992field}. The main features of this model can be reproduced by the simple and universal potential profile shown in Figure \ref{fig1}c where the triangular tunneling barrier is modified by the presence of a quantum dot potential. The down shifts of the peak are due to the poor screening of the external electric field by the quantum dot and are proportional to the distance between the dot and the metallic tip\cite{bennett1966theory}. 

\section*{Femtosecond excitation of an individual nano-object}
We studied several different individual \textit{in situ} fabricated quantum dots for different DC voltages and different laser intensities (from 0  to 320 GW/cm$^2$ nominal peak laser intensity, corresponding to 150 mW average laser power). The data presented in the following are from the same nano-object. Supporting Information I. H. shows results for a clean W tip with identical laser intensities, similar results on other quantum dots are shown in Supporting Information II and additional data analysis of this sample can be found in Supporting Information III but are not essential for the understanding of the photoemission mechanism.  

\begin{figure}
  \includegraphics[width=16cm]{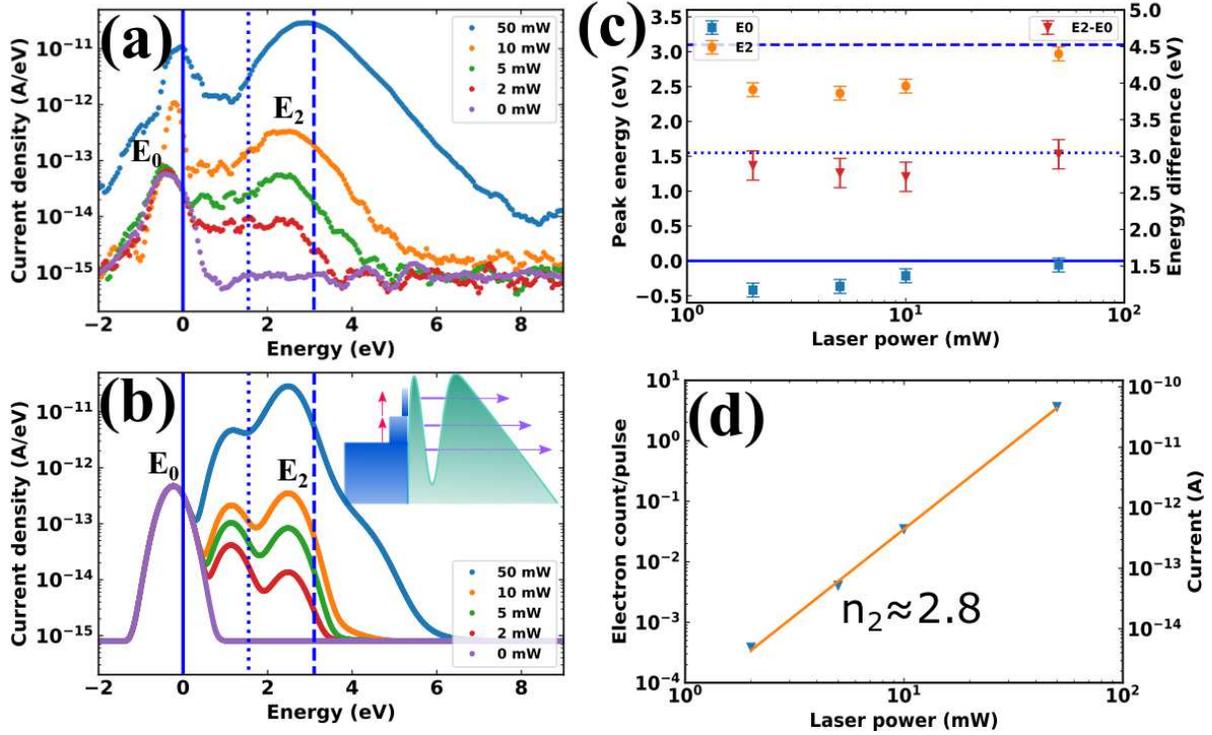}
  \caption{\label{fig2}(a) Experimental energy spectrum of the emitted electrons from a quantum dot for different femtosecond laser intensities (0, 4, 10, 21, 106 GW/cm$^2$) and a fixed applied voltage of 290 V. The vertical line represents the position of $E_F$ inside the W tip. The vertical dotted line indicates an emission energy $h\nu$ above $E_F$. The vertical dashed line indicates an emission energy $2h\nu$ above $E_F$. (b) Simulated energy spectrum of the emitted electrons from a quantum dot for different femtosecond laser powers. Inset: energy band diagram of the tunneling barrier for field emission with a quantum dot. (c) Energy $E_0$ (square) and $E_2$ (circle) of the peak maxima and their energy difference (triangle) as a function of the laser power. (d) Integrated current of the peak  $E_2$  as a function of the laser power. The solid line is a fit of the experimental data. n$_2$ is the slope of the linear fit}
\end{figure}

Figure \ref{fig2}a shows a typical electron energy spectrum as a function of the laser intensity for a fixed voltage which has a well-defined peak position $E_0$ at zero laser power. Upon excitation by the laser, an additional peak appears at an energy $E_2 = 2.5$ eV. It is important to notice that our measurements are a superposition of i) electrons continuously emitted because of the DC voltage and ii) laser excited electrons that are emitted during or shortly after the laser pulse. The $E_0$ peak is mostly emitted due to the DC field and it is difficult to estimate if its intensity changes come from the laser or from long term fluctuations. In contrast, the part of the spectrum above the Fermi energy comes from electrons emitted due to the laser pulse. Moreover, the instantaneous electron current at these energies is orders of magnitude larger than a comparable peak in the DC part of the spectrum. For a standard metallic tip, peaks at multiples of $h\nu$ above $E_F$ are expected\cite{schenk2010strong,yanagisawa2011energy}. Here the high energy peak position is clearly in between two expected values for laser powers below 50 mW (between $h\nu$ and $2h\nu$ above $E_F$). Figure \ref{fig2}c shows the evolution of the positions of the principal peaks for different laser powers. It can be noticed that the peak positions are rather constant and the energy difference between $E_2$ and $E_0$ is close to but somewhat less than $2h\nu$. The integrated current of the additional peak $E_2$ shown in Figure \ref{fig2}d has a power law dependence with laser power with an exponent of 2.8. This exponent can range from 1.5 to 4 depending on the applied voltage or nano-object studied.

Control experiments were carried out after removal of the nano-object by heating the tip at a temperature above 1000 K: i) it showed \textbf{no emission} at this DC voltage and laser power range ii) it recovered the emission characteristic of the tip before the nano-object growth. iii) the shape of the spectrum of the tip excited at an identical laser intensity as the quantum dot and identical total emitted electrons is different (see Figure S9b). This means that in the presence of the quantum dot, the extracted electrons travel only through the quantum dot, although the laser size and the tip area are much larger than the quantum dot. The field enhancement factor of the DC and laser field and the resonance of the quantum dot strongly counterbalance the small size of the nano-object.  

\begin{figure}
  \includegraphics[width=16cm]{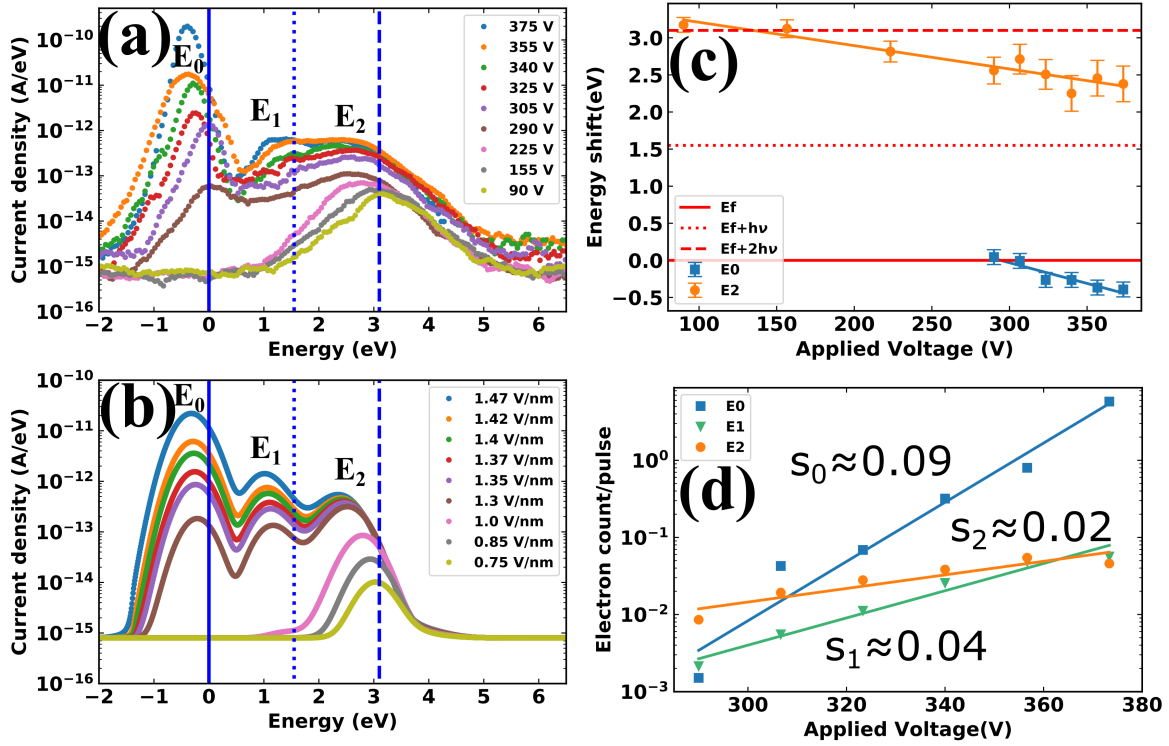}
  \caption{(a) Experimental energy spectrum of the emitted electrons from a quantum dot for different voltages at a fixed laser power of 10 mW (21 GW/cm$^2$). The vertical line represents the position of $E_F$ inside the W tip. The vertical dotted (respectively dashed) line indicates an emission energy $h\nu$ (respectively $2h\nu$) above $E_F$. (b) Simulated energy spectrum of the emitted electrons from a quantum dot for different fields. (c) Energy $E_0$ (square) and $E_2$ (circle) of the peaks maximum as the function of the applied voltage at a fixed laser power of 10 mW (21 GW/cm$^2$). The solid lines are a fit of the experimental data. (d) Integrated intensity of the three peaks as a function of the laser power. The solid line is a fit of the experimental data. s$_0$, s$_1$ and s$_2$ are respectively the slopes in V$^{-1}$ of the linear fit for $E_0$, $E_1$, and $E_2$.}
  \label{fig3}
\end{figure}

For low applied voltage (see in Figure \ref{fig3}a) a clear electron emission from $E_2$ is observed while the emission from $E_0$ cannot be detected. As the voltage is increased, emission from $E_0$ occurs and becomes dominant and another peak appears at an intermediate energy $E_1$. This intermediate peak is also in between two expected values (between $E_F$ and $E_F+h\nu$) strongly indicating that a different process takes place compared to W emitters. Compared to previous studies on photoemission of nanotips, our \textit{in situ} nanostructured photoemitter shows a drastically different dependence of the electron energy spectrum on DC voltage: for a fixed laser power, the peaks $E_0$, $E_2$ and probably $E_1$ shift linearly with DC voltage as shown in Figure \ref{fig3}c. In tungsten tips, it was only for the $E_2$ peak that some displacements have been observed\cite{yanagisawa2011energy} and this behavior originated from a mixing between the barrier lowering due to the Schottky effect and the $2h\nu$ peak. The $E_0$ and $E_1$ peak positions are always fixed for tungsten.  

The appearance of $E_0$ for a voltage above 290 V might seem intriguing but is simply explained and has already been observed in the past in the absence of laser excitation\cite{binh1992field}. At low voltage the $E_0$ state is weakly populated because it is above $E_F$. Its energy is too high to be sufficiently filled by electrons from the Fermi sea and too low to have a small tunnel barrier that allows promoted electrons by a photonic process to be emitted. The fact that the $E_1$ and $E_2$ peaks are not at the expected multiphotonics energies and shift with voltage indicate that these peaks come from the excited energy levels of the quantum dot as represented in the inset of Fig. \ref{fig2}b. The energy shift allows to tune the energy difference between electronic levels. 

Fig. \ref{fig3}d shows that the intensity of the three peaks has an exponential dependence with the applied voltage as expected for tunneling over a small voltage range. The slopes in Fig. \ref{fig3}d decrease with the order of the peaks in agreement with the idea that the higher the energy of a peak the lower the barrier height. Remarkably, despite the fact that the quantum dot is weakly bound to the nano-tip it maintains its quantized signature in the electron spectrum even upon laser excitation. Only minor instabilities related to flip-flop\cite{chen1979mobility} can be noticed above 10 mW and features can momentarily shift or even disappear in some measurements and automatically reappear. It indicates that the quantum dot can maintain its essential features even at high laser power and it does not lead to a sudden destruction of the tip as usually happens for unstable field emitters. As any clean field emitter in ultra-high vacuum, some modifications of the emitter occur on a time scale of an hour during which our experiments are performed. Fluctuations of the quantum dots can lead to slight displacement of the energy peaks less than 0.2 eV which can have a visible influence on the peak shapes close to the Fermi Energy as observed in ref.~\citenum{purcell1994field}. 
\section*{Interpretation of the photoemission process}

A first interpretation of the laser induced photoemission process would be to consider the direct photoexcitation of the quantum dot followed by electron emission. The measured value of $E_2-E_0$ that happens to be near $2h\nu$ could be interpreted as a 2-photon absorption process from the filled resonant energy level $E_0$ as observed for instance for Shockley surface states in Ref. \cite{reutzel2019coherent}. A first argument against the two photon excitation process can be made by examining the lower voltage curves in Figure \ref{fig3}a for which $E_0$ is not present. Thus no electrons are present to be excited to $E_2$ which is in contradiction with the fact that the $E_2$ peak is present in the spectra. To go further consider the time to refill the quantum dot energy level from the Fermi sea of the W tip. It is inversely proportional to the energy width of the peak ($<0.5$ eV) and hence is greater than 1 fs. The electron emission would not exceed roughly 10 electrons per pulse. Experimentally, we never observed a saturation of the current from $E_2$ in the high current regime as shown by the linear fit for the power law in the data in Figure \ref{fig2}d and in Supporting Information II A for another quantum dot at higher laser power Figure S.8. Moreover, the two photon cross section defined as $\sigma_2 = \frac{I_2 \tau}{ef} (\frac{h\nu f\pi w^2}{P})^2$, where $I_2$ is the integrated electron intensity for the peak centered at E$_2$, $\tau$ the pulse duration, $e$ is the electron charge, $f$ the laser repetition rate, $h\nu$ is the photon energy, $w$ the laser waist and $P$ the laser power, has a value in the $10^5$ GM range (Supporting Information III D). Such a high value can be observed in large quantum dots  and is orders of magnitude above the highest values reported in quantum dots of the same size\cite{chen2017size} (\textit{i.e.} between 1 and 2 nm as shown in the simulations below). This implies that the direct photo-excitation of an electron from the quantum dot is unlikely.

We propose a different mechanism which is that emission comes from electrons originally excited by the laser in the W tip. These electrons then preferentially tunnel through the excited states of the quantum dot to be emitted into the ionization continuum. As the tip is much larger than the quantum dot and has a much larger number of available electrons, a cross section several orders of magnitude larger than for a quantum dot is expected. This process is a combination of the resonant tunneling process observed in field emission\cite{lin1991observation,binh1992field} and the photoemission process observed in ref.\cite{yanagisawa2011energy}. In our case, it offers the possibility to have access to the electron dynamics from the tip to the quantum dot on an ultrafast timescale. 

\section*{Numerical simulations}
In order to confirm our hypothesis, numerical quantum calculations have been performed. The electron tunneling probability is obtained by solving the 1D Schr\"{o}dinger equation with the potential shown in Figure \ref{fig1}d. The independent parameters are the width of the first barrier, the size of the QD and the potential in the quantum dot. The parameters have been selected in order to reproduce the spacing and voltage dependence of the energy levels in the experiment. However we do not expect this model to predict the exact dimension or material of the quantum dot. We hope that these simple calculations, aimed at reproducing the main physical effects, will stimulate further more realistic calculations and experiments with different materials. The calculated transmission of the tunneling barrier is combined with the electron energy distribution inside the W tip to reproduce the emitted electron energy spectrum (see Supporting Information IV for details of the calculation). We calculated the time-dependent electron energy distribution in the out of equilibrium regime by including the effect of the laser pulse on the Boltzmann equation\cite{yanagisawa2011energy} of the electrons and the phonons in the metallic tip. 

The results of the simulations are presented in Figure \ref{fig2}b and \ref{fig3}b when the laser is on and in Supporting Information II b in the equilibrium case without laser. We found reasonable agreement with the experimental results. The $E_1$ peaks appear more clearly in the simulations than in the experiment because experimentally a peak is hardly detectable when another peak is present at higher energy with a higher intensity. The main reason is that the shot noise of the high energy peak overwhelms the signal coming from the low energy peak. It is also possible that additional scattering mechanisms not taken into account in our simulations might attenuate the amplitude of the $E_1$ peaks as well as make the $E_2$ peaks wider. Note that in our simulations the contribution of the laser field to the shift of the energy levels was not taken into account although in the range of power explored here it starts to be comparable to the DC field (10 mW corresponds to a laser field of 0.4 V/nm and an intensity of 21 GW/cm$^2$). It can be expected that with a higher intensity carrier envelop phase stabilized laser, the energy levels of the quantum dot might oscillate with laser electric field and present interesting new features. 

\section*{Discussion}

Compared to DC field emission, femtosecond laser excitation permits to create an ultrafast non-stationary electron distribution. The energy gained by the electrons from the photons allows the electrons to tunnel from the tip to the quantum dot higher energy levels. These states can therefore be observed in laser induced resonant tunneling. 
Because the laser pulse is short, the multiphoton absorption has to occur within a few fs and the non-stationary electron distribution is created at this timescale where electron-phonon and even electron-electron scattering is rather limited. In the Supporting Information 1c, our numerical simulations show that the out of equilibrium electron pulse in the metal at moderate laser intensity has a duration of $\sim 13$ fs, slightly smaller than the original laser pulse and this duration increases to 17 fs at 50 mW. Depending on the timescale of the tunneling process, the overall dynamics might be considered as a multiple step mechanism where excited electrons from the tip are created and have enough time to tunnel to the quantum dot and then into vacuum before the relaxation to the stationary Fermi-Dirac distribution and thermalization to the phonons occur. Future pump-probe experiments are expected to give further insight about the time domain and the energy range where the multiphoton process\cite{grubisic2012plasmonic} dominates over thermionic emission\cite{tan2017ultrafast} and the type of  coherent process involved such as the one identified in Ref.~\citenum{reutzel2019coherent}. 

This multi-step process has been unexplored so far in the ultrafast photoemission of nanotips. It is a promising approach that filters and favors the creation of coherent electrons from an artificial atom. In our case, from the peak width we can estimate that the resonant tunnelling process occurs within a few fs and with a pulse duration of 14 fs the multistep scenario is fulfilled. A gain of a factor of 10 is still possible in order to obtain a transform limited electron pulse. Such an improvement is within reach either by increasing the width of the first barrier by few \AA ngstr\"oms (a 64 meV peak width on the field emission of a quantum dot has already been reported in the past\cite{purcell199564}) or by reducing the laser pulse duration.

\section*{Conclusions}

We have performed field assisted photoemission induced by an ultrashort laser pulse on a single, isolated quantum dot attached to a metallic tip. We demonstrated that the emission process in the range of nominal peak laser intensity studied here (up to 320 $GW/cm^2$) is well-described by a multiphotonic process where electrons in the metal tip tunnels resonantly through the quantum dot as shown by numerical simulations. The presence of an additional DC voltage offers the possibility to finely tune and control the energy difference between electron energy levels for interference experiments. This tunneling process on tunable and well defined atomic-like states paves the way for the development of high brightness, ultra-fast and coherent single electron sources for quantum optics studies. Measurements of the electron spatial distribution and energy levels spectroscopy of the quantum dot with a tunable laser\cite{mingels2015sensitive} could give deeper understanding of the 3D electron trajectories and their full interaction with the laser. To finish, now that we have established access to the femtosecond laser excitation of quantum dot states, note that other quantum nanostructures\cite{kleshch2020carbon} or even 1D objects\cite{pascale2014ultrashort} may emit from their quantum states by other mechanisms such as direct photo excitation, thus opening up other avenues of investigation.

\begin{acknowledgement}

This work was supported by the ILM AAP project.
The authors acknowledge the Plateforme Nanofils et Nanotubes Lyonnaise of the University Lyon1. H.Y. and M.F.K. are grateful for support by the German Research Foundation (DFG) via YA-514/1-1 and the Munich Centre of Advanced Photonics. M.F.K. acknowledges support by the Max Planck Society through the Max Planck Fellow program.

\end{acknowledgement}

\begin{suppinfo}

This material is available free of charge via the internet at http://pubs.acs.org. Additional description of the experiment Additional data and analysis. detailed presentation of the model and numerical simulations.

\end{suppinfo}

\bibliography{nanop2}

\end{document}